\documentclass[twocolumn,showpacs,amsmath,amssymb]{revtex4}
\usepackage{amsmath}
\usepackage{amssymb}
\begin{document}
\baselineskip 15pt
\title{\bf Statistical Analysis of Airport Network of China
\footnote{Supported in part by National Natural Science Foundation of China under No. 70271067, a key
program of Hua-Zhong Normal University, and a grant from the international program of the Santa Fe
Institute of the United States of America.}}

\author{W. Li$^{1,2,3}$ and X. Cai$^1$}
\affiliation{$^1$Institute of Particle Physics, Hua-Zhong
Normal University, Wuhan 430079, China \\$^2$Max-Planck Institute for Mathematics in the Sciences,
Inselstr. 22-26,
D-04103 Leipzig, Germany \\
\rm $^3$Electronic Address: liwei@mis.mpg.de}
\date{August 12, 2003}

\begin{abstract}
{\em Through the study of airport network of China (ANC), composed
of 128 airports (nodes) and 1165 flights (edges), we show the
topological structure of ANC conveys two characteristics of small
worlds, a short average path length (2.067) and a high degree of
clustering (0.733). The cumulative degree distributions of both
directed and undirected ANC obey two-regime power laws with
different exponents, i.e., the so-called Double Pareto Law.
In-degrees and out-degrees of each airport have positive
correlations, whereas the undirected degrees of adjacent airports
have significant linear anticorrelations. It is demonstrated both
weekly and daily cumulative distributions of flight weights
(frequencies) of ANC have power-law tails. Besides, the
weight of any given flight is proportional to the degrees of both airports at
the two ends of that flight. It is also shown the diameter of each
sub-cluster (consisting of an airport and all those airports to
which it is linked) is inversely proportional to its density of
connectivity. Efficiency of ANC and of its sub-clusters are measured through
a simple definition. In terms of that, the efficiency of
ANC's sub-clusters increases as the density of connectivity does.
ANC is found to have an efficiency of 0.484.}
\end{abstract}

 \maketitle The ER model
\cite{ER} of random graphs, introduced by Erd\"{o}s and R\'{e}nyi,
has attracted much attention from both mathematicians and
physicists \cite{Bollobas, Mendes, Albert-Barabasi-1, Ben-Havlin,
Christensen}, and henceforth shaped our understanding of networks
for decades. The growing interest on whether randomness dominates
real-world networks, however, was eventually prompted by recent
advances in two main streams of topics. One part of these work
was related to ``small worlds'', originally described as ``six
degrees of separation'' \cite{Milgram} which manifests that humans
are connected through a short, limited chain of acquaintances. The
concept was successfully employed by Watts and Strogatz \cite{WS, Watts}
in exploring the dynamics of a great variety of networks between
order and randomness, e.g., the actor and actress networks
\cite{MovieActor}, the chemical reaction networks
\cite{ChemicalReaction}, the rumor spreading networks
\cite{Watts}, and the food webs \cite{FoodWebs}, etc. Another
parallel achievement was made by the research team of
Barab\'{a}si \cite{Albert-Barabasi-2, Albert-Jeong-Barabasi,
Jeong, Yook}, which led to the finding of a class of networks with
scale-free degree distributions, for example, Internet
\cite{Albert-Barabasi-2}, the networks of co-authorship in natural
sciences \cite{Newman1}, the web of sexual contacts
\cite{Liljeros}, and the graph of human language
\cite{Ferrer-Sole}, etc.

Composed of a number of airports and flights, air networks are
simply normal examples of transportation systems among ubiquitous networks in nature.
Nevertheless, they appear extraordinary and unique due to the
following features: (a) quite limited system sizes, from a few
hundred to a few thousand at most; (b) relatively stationary structures with respect to both time
and space; (c) bi-directional, weighted links (flights) with
slightly fluctuating frequency.

This paper will present
investigations of airport network in China (ANC). We demonstrate that
on one hand ANC embodies part features of small worlds and of
scale-free networks. On the other hand, however, ANC exhibits more features
belonging to its topological uniqueness. The whole text is
organized as follows. Section I presents the results on degree
distributions and degree correlations of ANC. Section II gives the flight
weight distributions and the weight-degree correlation of ANC.
Section III analyzes the clustering coefficients of ANC.
In Section IV we calculate the diameter of ANC and discuss the
efficiency of ANC by proposing a simple definition for it. Conclusions and discussions
are given in the last part, section V.

\section {Degree distributions and degree correlations}

ANC consists of $N=128$ \cite{note1} airports (nodes) and 1165
flights (edges) that connect most major cities in China. The
topology of ANC can be symbolized by a $128 \times 128 \times 7$
connectivity matrix $C$ whose entry $C_{ijt}$ is 1 if their is a
link pointing from node $i$ to node $j$ at the $t$-$th$ day of a
week (Herein and after $t$=1, 2, 3, 4, 5, 6, and 7 specifies the
seven days within a week, starting from Monday, respectively.) and
0 otherwise, and a $128 \times 128 \times 7$ weight matrix $W$
\cite{note2} whose element is defined as
\begin{equation}
\label{weight}
W_{ijt}=\frac {n_{ijt}} {\sum_{t}\sum_{\{i,j\}} n_{ijt}},
\end{equation}
where $n_{ijt}$ is the number of flights $i \rightarrow j$ at the $t$-$th$ day. $W_{ijt}$
satisfies the normalization condition, i.e., $\sum_{t}\sum_{\{i,j\}} W_{ijt}=1$.
Normally, $C_{ijt}=C_{jit}$ and $W_{ijt}=W_{jit}$ only hold for undirected ANC.

We employ $k_{in}^{w}(i)$ and $k_{ou}^{w}(i)$ to denote the in-degree and out-degree of a given node $i$ in the
directed ANC during a whole week time, and $k_{un}^{w}(i)$ to represent the undirected degree of the
undirected ANC in the same week.
Hence, we have
\begin{equation}
\label{in-degree}
k_{in}^{w}(i)=\sum_{j}^{j \neq i} \eta (\sum_{t=1}^{7} C_{ijt} -1),
\end{equation}
\begin{equation}
\label{out-degree}
k_{ou}^{w}(i)=\sum_{j}^{j \neq i} \eta (\sum_{t=1}^{7} C_{jit} -1),
\end{equation}
\noindent and
\begin{equation}
\label{all-degree}
k_{un}^{w}(i)=\sum_{j}^{j \neq i} \eta (\sum_{t=1}^{7} [C_{ijt}+C_{jit}] -1),
\end{equation}
\noindent
where $\eta (x)$ is a unit step function, which takes 1 for $x \geq 0$ and 0 otherwise.

First we consider the three distributions of $k_{in}^{w}(i)$, $k_{ou}^{w}(i)$, and $k_{un}^{w}(i)$,
respectively.
Here the cumulative distribution, widely used in economies and well known as the Pareto Law \cite{Pareto1},
is adopted to reduce the statistical errors arisen from the limited system size. The cumulative
form, $P (k_{in}^{w}(i) > k)$ ($P (k_{ou}^{w}(i) > k)$, or $P (k_{un}^{w}(i) > k)$), gives the probability
that a given airport $i$ has an in-degree (out-degree or undirected degree) larger than $k$. Fig. 1A presents
behaviors of the three distributions. It is amazing to find that all three distributions follow nearly
a same two-regime power law with
two different exponents, known as {\sl Double Pareto Law} \cite{Double_Pareto}, with a turning point
at degree value $k_c \simeq 26$, which can be well prescribed by the following expression,
\begin{equation}
\label{e1}
 P(K > k) \thicksim
\begin{cases}
k^{-\gamma_1},&\text{for $k \leq k_c$}; \\
k^{-\gamma_2},&\text{for $k > k_c$},
\end{cases}
\end{equation}

\noindent where $\gamma_1$ and $\gamma_2$ are the respective
degree exponents of two separate power laws. By means of fitting,
exponents pairs ($\gamma_1$, $\gamma_2$) of the three
distributions in Fig. 1A are (0.428, 4.161), (0.416, 4.453), and
(0.45, 4.535). Using a simple algebra, the original distributions
of $k_{in}^{w}(i)$ ($k_{ou}^{w}(i)$, or $k_{un}^{w}(i)$) can be
written as,
\begin{equation}
\label{original-degree} P(k)=\frac {\partial P(K > k)}{\partial k}
\thicksim
\begin{cases}
k^{-(\gamma_1+1)},&\text{for $k \leq k_c$}; \\
k^{-(\gamma_2+1)},&\text{for $k > k_c$},
\end{cases}
\end{equation}
\noindent where $k$ specifies the three different degrees above.
Correspondingly, the mean values of $k_{in}^{w}(i)$, $k_{ou}^{w}(i)$,
and $k_{un}^{w}(i)$ are 18.931, 17.156 and 18.203. This conveys that
each airport, on average, is connected to around 18 other airports.

The undirected degree of a certain airport $i$ at the $t$-$th$ day of a week is given by
\begin{equation}
\label{undirected}
k_{un}^{t}=\sum_{j}^{j \neq i} \eta (C_{ijt}+C_{jit}-1).
\end{equation}
\noindent The cumulative distributions of $k_{un}^{t}$, with
$t=$1, 2, 3, 4, 5, 6, and 7, shown in Fig. 1B, reflects the daily
evolution of the topology of the undirected ANC within a week. It is evident from
Fig. 1B that the distributions of days from Monday to Saturday
nearly coincide with one another, on a same Double Pareto Law. The
distribution of Sunday, however, deviates apparently from the
shared curve and itself obeys another law. By checking the
original data, one may find out the discrepancy is mainly caused
by the fact that some flights are not available on Sundays.
Exponents pairs and average undirected degrees of the undirected ANC for each day of one week
are listed in Table I. As we can see, the values of $\gamma_1$ and $\gamma_2$ in the table are
in general (except on Sundays) slightly larger than the counterparts of undirected ANC during
a whole week. The average degrees of each day, around 14 (12 on Sundays), are much smaller
than 18, the counterpart of a week. This is understandable because many flights are
only available on certain days of a week.

We also check an important feature of ANC, the degree
correlations. First we come to the correlation between in-degrees
and out-degrees, simply called in-out degree correlation. This is quite natural for airport networks
because each airport should generally maintain the balance of its
traffic flow. Normally, for each airport, the higher its
in-degree, the higher its out-degree. We plot $k_{in}^{w}(i)$
versus $k_{ou}^{w}(i)$ (i=1,2,...,128) in Fig. 2. The following
expression can be obviously obtained by fitting the data,
\begin{equation}
\label{in-out-correlation}
 k_{in}^{w}(i) \simeq k_{ou}^{w}(i).
\end{equation}
\noindent Evidently, the in-out degree correlation is very strong.

Another possible correlation exists
between the degrees of adjacent airports, named degree-degree correlation. The degree-degree correlation
tells that the degrees are not independent and correlate with those of their neighbors.
It can be demonstrated by calculating
the mean degree of the neighbors of a given airport as a function
of the degree of that airport. Fig. 3 presents our analysis of
degree-degree correlation in the undirected ANC. As shown, the
degrees of adjacent airports have significant anticorrelations,
based on which the ANC appear to be disassortative
\cite{disassortative-network}. But the anticorrelation
found in ANC is almost linear, different than that found in Ref. \cite{anti-degree-correlation},
which is a power law with exponent of about -0.5.

\section {Flight weight distributions and weight-degree correlation}

An important feature of ANC is that some flights are more frequent
than others. The weight or the frequency of a certain flight, henceforth, states the extent to which
it is busy. The weight of flight $i \rightarrow j$ in a whole
week is given by
\begin{equation}
\label{weight1}
W_{ij}^{w}=\sum_{t=1}^{7} W_{ijt}.
\end{equation}
\noindent
The cumulative distribution of $W_{ij}^{w}$, $P (W_{ij}^{w} > W)$, gives the probability that a flight
has a weight larger than $w$. Shown in Fig. 1C, $P (W_{ij}^{w} > W)$ has a power-law tail,
\begin{equation}
\label{e2}
P (W_{ij}^{w} > W) \thicksim W^{-\gamma},
\end{equation}
\noindent
where $\gamma=1.65$. Through a simple algebra, one may obtain $P (W) \thicksim (W)^{-2.65}$.
Such a power-law tail indicates that the probability of finding a very busy flight is nonzero, and significant
instead. The daily cumulative distributions of $W_{ijt}$ within a week is given in Fig. 1D.
Amongst the seven distributions, those from Monday to Saturday obey the same power law, while
Sunday data reveals a steeper power law that extends a narrower region on the x-coordinate. The exponents of
flight weight distributions of each day are also presented in Table I and are slightly larger than 1.65.

We also conjecture if there is a certain kind of relation between the weight of a given flight
and the degrees of the two airports at both ends of that flight. We simply call it weight-degree correlation.
Without losing the generality, we propose the following ansatz for the possible existence of such correlation,
\begin{equation}\label{weight-degrees}
W_{ij}^{w} \thicksim [k_{un}^{w}(i)*k_{un}^{w}(j)]^{1/2}.
\end{equation}
\noindent
This scaling ansatz has been well supported by the real data, shown in Fig. 4.

\section {Clustering coefficient}

The neighborhood $\Gamma_v$ of a given airport $v$ is a graph
which includes all nodes who have flights with $v$. The clustering
coefficient \cite {Watts} $C(\Gamma_v)$ of neighborhood $\Gamma_v$
of airport $v$ characterizes the extent to which airports in
$\Gamma_v$ are connected to every other. In precise words,
\begin{equation}
\label{e3}
C(\Gamma_v)=\frac {E(\Gamma_v)} {C_m^2},
\end{equation}

\noindent
where $E(\Gamma_v)$ is the number of real connections in $\Gamma_v$ consisting of $m$ airports,
and $C_m^2$ is the total number of all possible connections in $\Gamma_v$. The average clustering coefficient
of the entire air network is defined as,
\begin{equation}
\label{averageclustering}
C=\frac{1}{N}\sum_{\Gamma_v} C(\Gamma_v),
\end{equation}
\noindent
where $N$ is the number of airports of the whole network. By calculation, $C$ of the entire undirected
ANC for a whole week is 0.733, in stark contrast with the low density of connectivity,
$\langle k \rangle /N$, 0.143. The $C$ of the daily undirected ANC given in Table I centralizes 0.600,
the value for Sunday being slightly lower.

\section {Diameter and efficiency}

For a connected network, the diameter $D$ can have the following
definition,
\begin{equation}
\label{diameter}
D=\frac{1}{N(N-1)/2} \sum_{(i,j)} d_{min}(i,j),
\end{equation}
\noindent
where $d_{min}(i,j)$ represents the shortest-path length between nodes $i$ and $j$.
In an air network, the diameter $D$ indicates the average number of transfers a passenger need to
take between the start and the end. For ANC, $D$ is around 2.067. Specifically, $d_{min}(i,j)$
in ANC only takes three distinctive values, 1, 2 and 3, with percentages of 0.143, 0.646, and 0.211,
respectively. This implies most trips will need one intermediate transfer or two before the final
destinations, only a small percent can be reached directly.

The high clustering and the small diameter inevitably indicates the small-world property of ANC.
For comparison, random graphs of the
same average degree, $\langle k \rangle$, and the same number of nodes, $N$, with ANC are investigated.
It is readily to learn that the average clustering coefficient of random graphs, 0.143, is
much smaller than 0.733, the weekly average clustering coefficient. The diameter of such random graphs,
scales as $\ln N/ \ln \langle k \rangle$, which is 1.672, less than the counterpart of ANC.

A practical thing of ANC is related to its transportation efficiency, which tells us how one can travel
from one place to another both quickly and economically. Let us first take a look at the efficiency of
sub-clusters of ANC. A sub-cluster here is composed of
a hub $v$, the central node, and its neighborhood $\Gamma(v)$ consisting of whoever has flights
with the hub. The largest sub-cluster of ANC includes 84 airports, and the smallest one, only 2.
In terms of graph theory, the sub-clusters consist of only two kinds of structure, trees and triangles.
The density of connectivity of a sub-cluster with $m$ nodes and $E(\Gamma_v)$ edges in $\Gamma(v)$ of
the hub is,
\begin{equation}
\label{e4}
\rho_{dc}=\frac {2(E(\Gamma_v)+m)}{m^2+m}.
\end{equation}

\noindent
The diameter of the
sub-cluster, $d_{sc}$, can be readily derived,
\begin{equation}
\label{e5}
D_{sc}=\frac{2(m^2-E(\Gamma_v))} {m^2+m}.
\end{equation}
\noindent
The plot of $D_{sc}$ versus $\rho_{dc}$, for all 128 sub-clusters of ANC, is presented in Fig. 5A,
which can be well fitted by a straight line. The larger $\rho_{dc}$ is, the more direct connections there
exist in the sub-clusters, and the smaller the diameter will be. In the case of a complete graph,
the diameter will be definitely 1.

We simply define the efficiency of sub-clusters of ANC,
$E_{sc}$, as,
\begin{equation}
\label{e6}
E_{sc}=\frac {1} {D_{sc}}=\frac {m^2+m}{2(m^2-E(\Gamma_v))}.
\end{equation}
\noindent
After a simple calculation, $E_{sc}$ versus $\rho_{sc}$ is presented in Fig. 5B. It's clearly shown that
the higher the density of connectivity, the higher the efficiency of a sub-cluster. The efficiency is
1 when the sub-cluster is totally connected. This agrees with our intuition.

Compared with its sub-clusters, ANC itself displays no more difference in structure.
The ANC can be viewed as a cluster with hierarchical structure \cite{Li_Cai}, composed of a center,
e.g., Beijing, and whoever
has direct connections with the center, and whoever has no direct connections with the center, but with
whoever has, and so on. For a connected network, such a cluster can include all nodes in the same system.
By analyzing the real data,
each node of ANC is connected to any other with no more than three steps. In this sense, Eq. (\ref{e6})
also applies to ANC. After some algebra, we find the efficiency of ANC is 0.484.

\section {Conclusions and discussions}

In conclusion, our analysis reveals two characteristic small-world
properties of ANC, a short average path length and a high degree
of clustering. Another important feature of ANC, the degree
distribution, however, is strikingly different from counterparts
of both scale-free networks and of random graphs. In ANC there exist strong, positive
correlations between in-degrees and out-degrees of each airport, and significant anticorrelations
between degrees of adjacent airports. The weekly and
daily weight distributions of ANC display power-law behaviors. The existence of weight-degree correlation
of ANC shows that there is an dependence of the weight of a certain flight on the degrees of the two airports
at both ends of that flight. In
particular, we suggest a rough idea to measure the efficiency of
ANC and that of its sub-clusters.

In the previous sections we do not answer why the structure of
ANC obeys double Pareto Law. Here we come up with a simple idea which can be
realized through computer simulation.
Suppose one constructs a whole
airport network from the very beginning, with only a few airports
in major cities, following two simple rules. Under the first rule,
preferential attachment \cite{Albert-Barabasi-2}, a newly
established airport tends to connect to the hubs with more
flights, which naturally drives the airport network to develop a
structure beyond those of random graphs. The second rule manifests
the existence of different growth rates of airports between the
region of smaller airports and that of larger ones. Thats is, in the early
history of airport network construction, smaller airports have
considerable probabilities to be growing to accommodate more
flights. Gradually, as most major airports have been established,
the smaller airports were unlikely to expand any more. Hence more
small-sized airports were established. This limited growth endows
the airport network features part of scale-free topology. It may
be more appropriate to say that ANC has an intermediate topology
between random graphs and scale-free networks.

Another issue should be addressed to the efficiency. The efficiency based on
our definition is solely idealistic and only limited to the structure of the network itself. It is more
appropriate to call it structural efficiency.
In the reality of air transportation, the carriers (airlines) should consider more factors
in order to have a higher and reasonable efficiency. That is, one needs to know how an
air network can satisfy the passengers' needs on one hand, and ensure the profits of airlines,
on the other hand. This should be an interesting topic and worth investigating.

W.L. would like to thank Alexander von Humboldt Stiftung for research funding and Prof. Juergen Jost
of Max-Planck Institute for Mathematics in the Sciences for hosting.

\begin{table}
\caption{}
  \begin{tabular}{cccccccc} \hline\hline
               & Mon & Tue & Wed & Thu & Fri & Sat & Sun  \\
   \hline
   $\gamma_1$  &0.582&0.569&0.568&0.603&0.558&0.574&0.463 \\
   \hline
   $\gamma_2$  &4.398&4.190&4.338&3.949&4.308&4.264&3.992 \\
   \hline
   $\langle k \rangle$ &13.570 & 14.376& 13.967& 14.017& 14.033& 14.586& 12.264 \\
   \hline
   $\gamma$ &1.744&1.682&1.729&1.699&1.679&1.747&2.329 \\
   \hline
     $C$    &0.626&0.621&0.614&0.590&0.638&0.620&0.576 \\
     \hline\hline
  \end{tabular}
\end{table}

\vskip 0.2cm
\begin{center}
\bf Figure Captions:
\end{center}

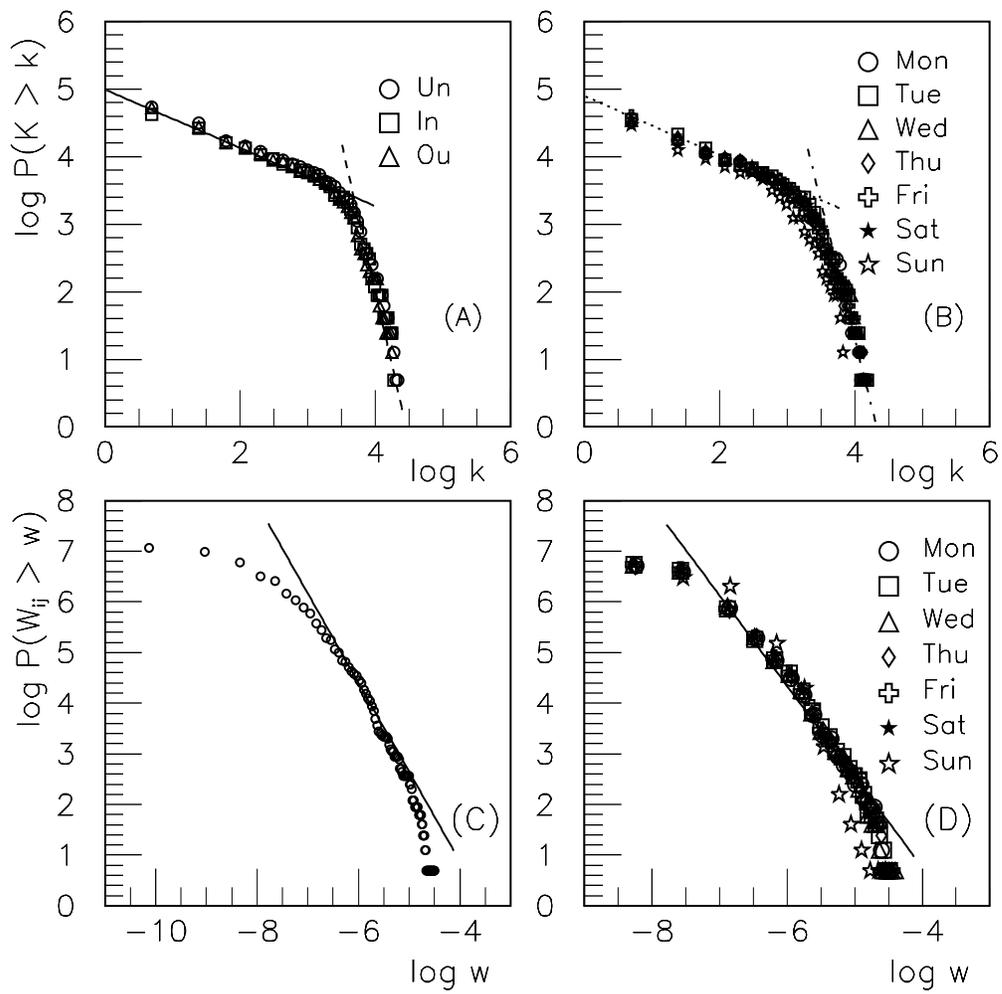
Fig. 1: Cumulative degree distributions of ANC for undirected degree, in-degree, and out-degree of (A) a whole week and (B) each day
from Monday to Sunday. Cumulative weight distributions of ANC for (C) a whole week and (D) each day
from Monday to Sunday.

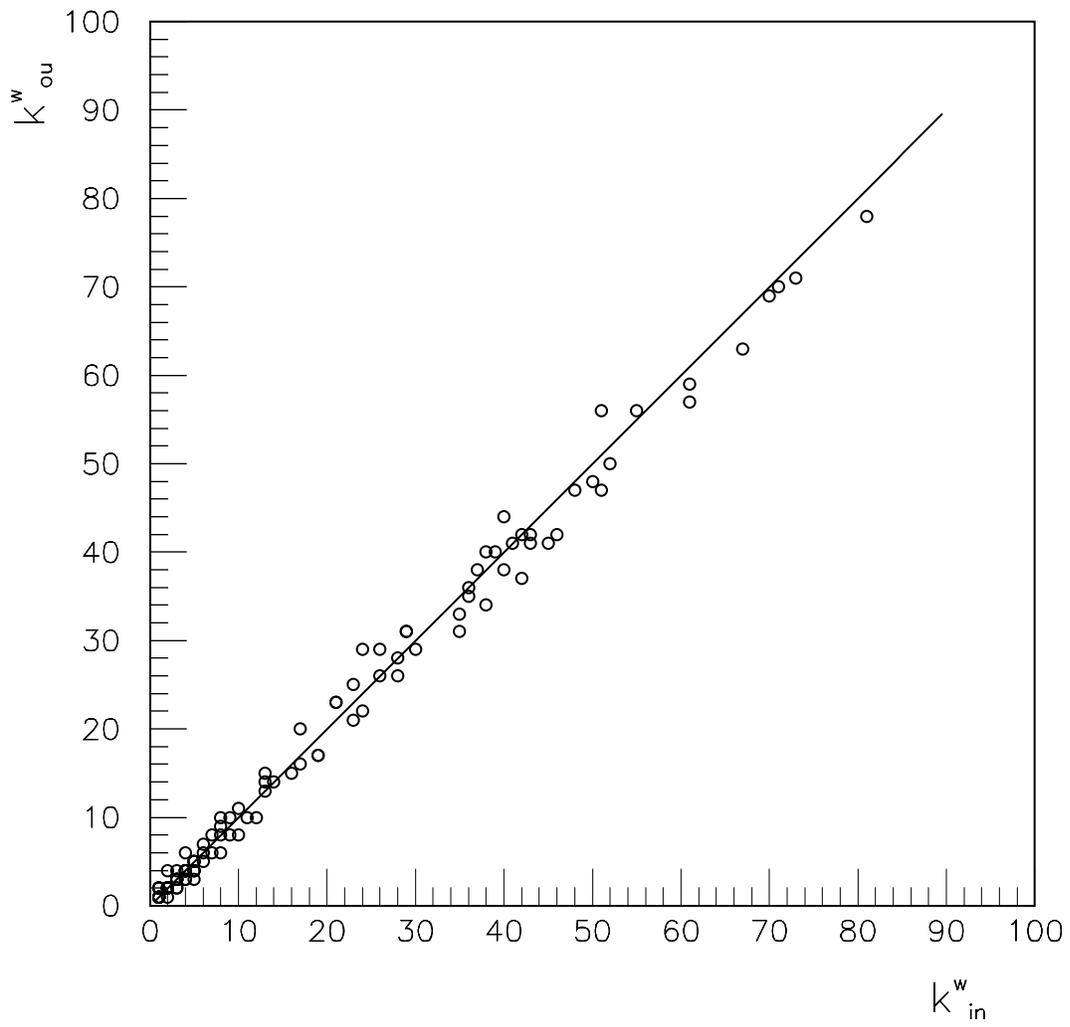
Fig. 2: Correlation between in-degrees and out-degrees of the directed ANC in a whole week.

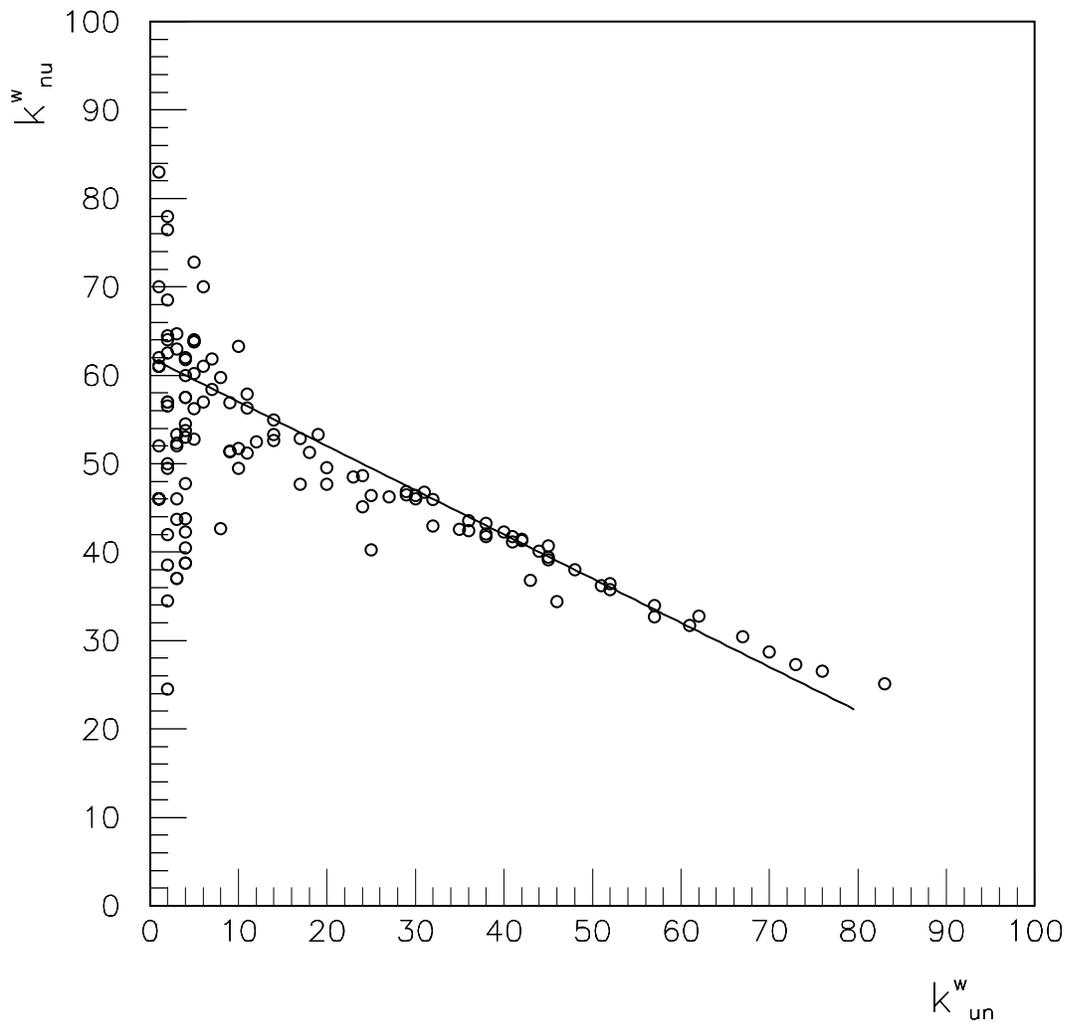
Fig. 3: Correlation between degrees of adjacent airports of the undirected ANC in a whole week.

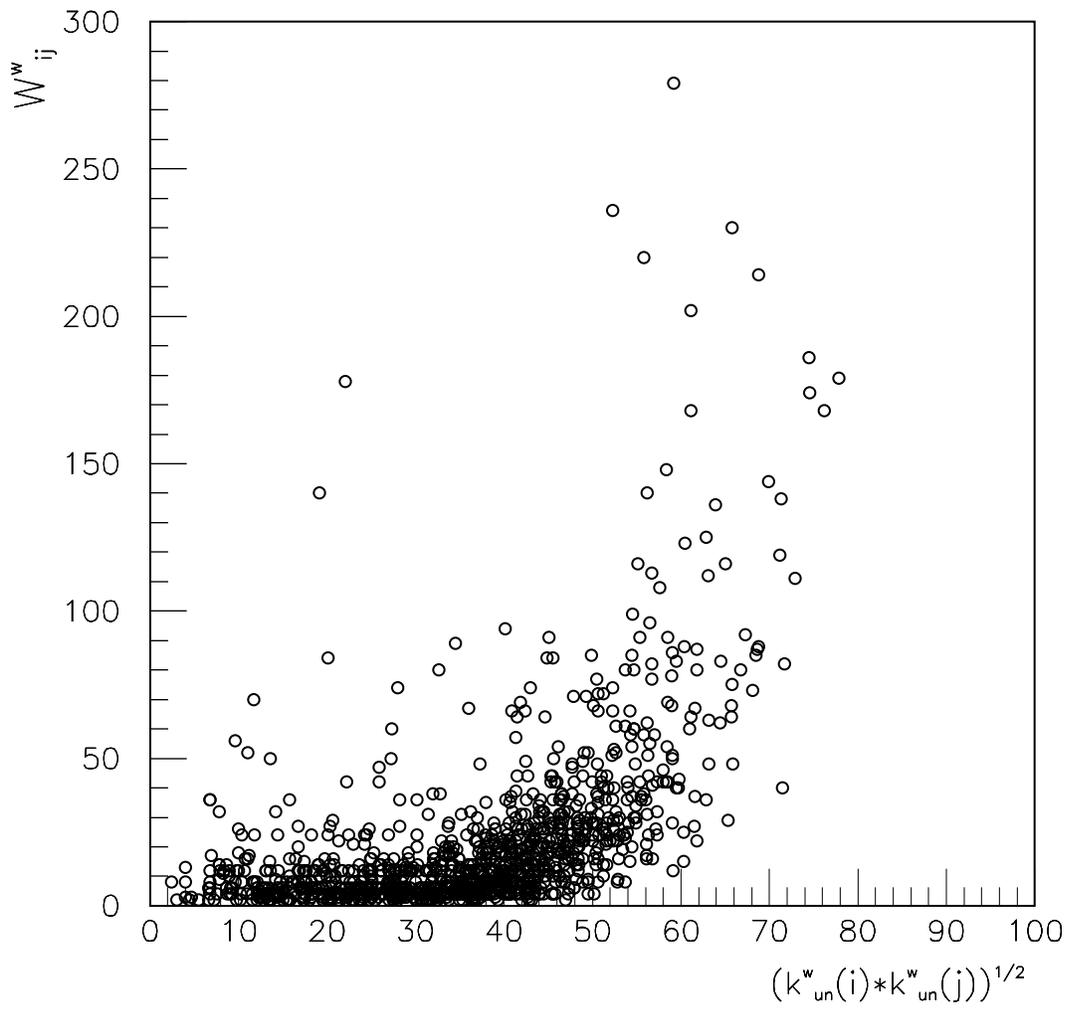
Fig. 4: Weight-degree correlation of the undirected ANC in a whole week.

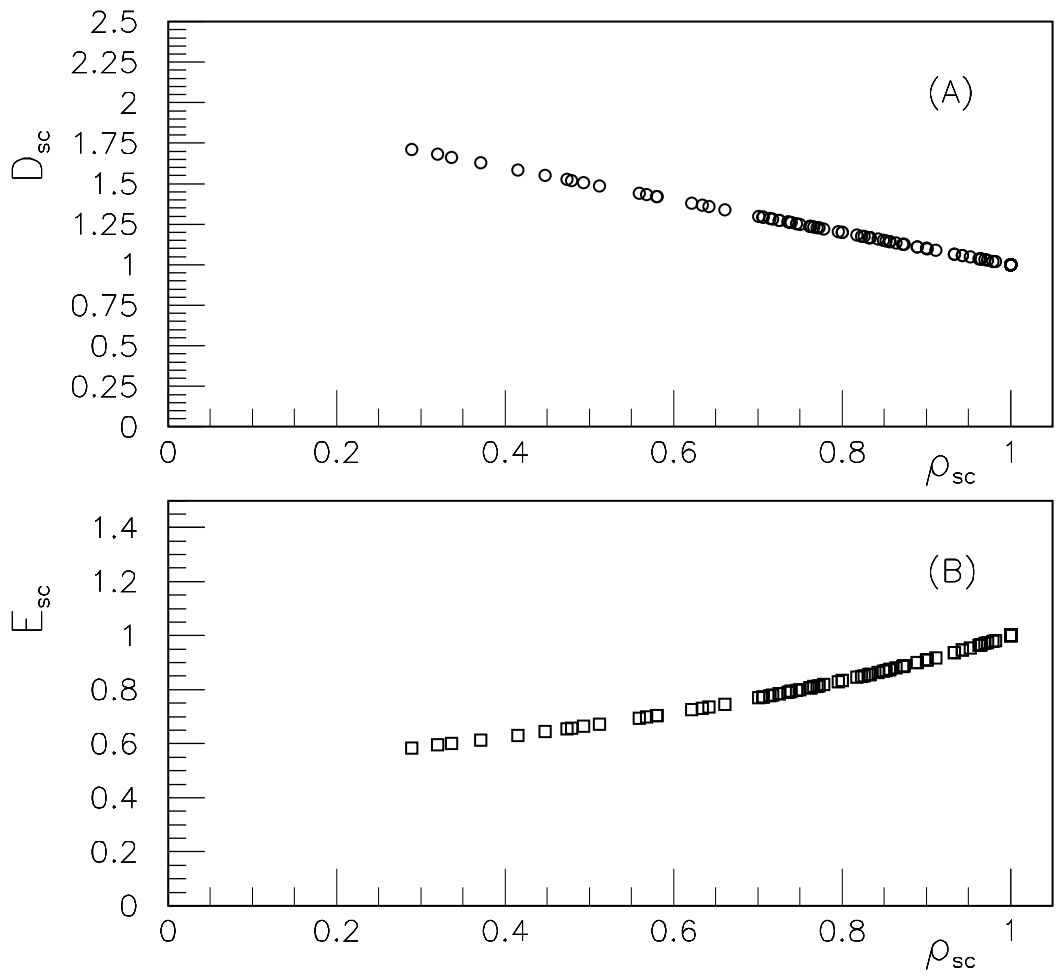
Fig. 5: (A) Diameter and (B) Efficiency versus density of connectivity for sub-clusters of the undirected ANC.

\vskip 0.2cm
\begin{center}
\bf Table Captions:
\end{center}

TABLE I: Comparison of relevant variables of daily undirected ANC (from Monday to Sunday):
(1) $\gamma_1$ and (2) $\gamma_2$ are exponents of two power-laws of cumulative degree distributions;
(3) $\langle k \rangle$, the average degree; (4) $\gamma$, the exponent
of flight weight distributions; (5) C, the clustering coefficient of the whole system.


\begin{thebibliography}{s2}
\bibitem{note1} The data was taken from a web site. According to it, the total number of airports in China
is 129, within which Pudong and Hongqiao belong to the same city Shanghai and are reagrded as united.
\bibitem{note2} Strictly speaking, the air network should also include a matrix of spatial distance $L$.
In this paper, $L$ is not taken into account.
\end{thebibliography}
\end{document}